\documentclass{an}
\usepackage{times}
\usepackage{graphicx}
\graphicspath{{./fig/}{./png/}}
\usepackage{fancyhdr}
\usepackage{bm}
\newcommand{\EQ}{\begin{equation}}
\newcommand{\EN}{\end{equation}}
\newcommand{\EQA}{\begin{eqnarray}}
\newcommand{\ENA}{\end{eqnarray}}

\newcommand{\bra}[1]{\langle #1\rangle}

\def\ga{\mathrel{\mathchoice {\vcenter{\offinterlineskip\halign{\hfil
$\displaystyle##$\hfil\cr>\cr\sim\cr}}}
{\vcenter{\offinterlineskip\halign{\hfil$\textstyle##$\hfil\cr>\cr\sim\cr}}}
{\vcenter{\offinterlineskip\halign{\hfil$\scriptstyle##$\hfil\cr>\cr\sim\cr}}}
{\vcenter{\offinterlineskip\halign{\hfil$\scriptscriptstyle##$\hfil\cr>\cr\sim\cr}}}}}
\title{Primordial magnetic fields and CMB anisotropies}
\author{Kandaswamy Subramanian}
\institute{
IUCAA, Post Bag 4, Pune University Campus, Ganeshkhind, Pune 411 007, India
}

\date{}
\begin{document}

\abstract{Possible signatures of primordial magnetic fields
on the Cosmic Microwave Background (CMB) temperature and polarization
anisotropies are reviewed. The signals that could be searched for include
excess temperature anisotropies particularly at small angular scales 
below the Silk damping scale, B-mode polarization, and non-Gaussian
statistics. A field at a few nG level produces temperature anisotropies
at the $5 \mu$K level, and B-mode polarization anisotropies 10 times smaller, 
and is therefore potentially detectable via the CMB anisotropies.
An even smaller field, with $B_0 < 0.1$ nG,
could lead to structure formation at high redshift $z > 15$, and hence
naturally explain an early re-ionization of the Universe. 
\keywords{cosmic microwave background -- cosmology: theory -- magnetic fields}}

\maketitle

\section{Introduction}

Magnetic fields are ubiquitous in astrophysical systems but
their origin is still a mystery. One possibility is that they
arise due to the dynamo amplification of small seed fields.
The dynamo paradigm has been extensively studied (cf. Brandenburg 
\& Subramanian 2005 for a recent review), but potential
problems remain to be surmounted. These involve the question of
whether mean field dynamo coefficients for the galactic dynamo
are catastrophically quenched or not, and whether fields generated
by the fluctuation dynamo in clusters have a 
sufficient degree of coherence to explain the observations
(Subramanian, Shukurov \& Haugen 2006; Shukurov, Subramanian \& Haugen
2006; Schekochihin et al. 2005).
 
An interesting alternative is that the observed large-scale 
magnetic fields are a relic from the
early Universe, arising perhaps during inflation or some other
phase transition (Turner \& Widrow 1988; Ratra 1992; and reviews by 
Widrow 2002; Giovannini 2005a). 
It is well known that scalar (density or potential)
perturbations and gravitational waves (or tensor perturbations)
can be generated during inflation. Could
magnetic field perturbations could also be generated? 
Indeed mechanisms involving, say, string theory, for the
generation of primordial fields have been reviewed in this meeting
(Gasperini 2006 and references therein). These generically involve 
the breaking of
the conformal invariance of the electromagnetic action, and the
predicted amplitudes are rather model dependent. 
Nevertheless, if a primordial magnetic field 
with a present-day strength of $B \sim 10^{-9}$~G 
and coherent on Mpc scales is generated,
it can strongly influence Galaxy formation. An even weaker field, 
sheared and amplified due to flux freezing, 
during galaxy and cluster formation may ease the problems of
the dynamo. It is then worth considering if one can detect or
constrain such primordial fields. 

Here we review possible probes of primordial magnetic fields, 
which use CMB temperature and polarization anisotropies.
Indeed there are also a number of puzzling features associated
with current CMB observations, which are not fully understood.
These include (a) the excess power in temperature anisotropies detected
by the Cosmic Background Imager (CBI) experiment on small
angular scales (Readhead et al. 2004),
(b) the observations by the Wilkinson Microwave Anisotropy Probe (WMAP)
satellite for an unexpectedly high redshift of re-ionization (Kogut et al. 2003),
(c) the low CMB quadrupole anisotropy seen compared to theoretical
expectations (cf. Spergel et al. 2003), 
and (d) various asymmetry and alignment effects 
(cf. de Oliveira-Costa et al. 2004).
Such features also encourage us to keep open possibilities
of new physical effects, perhaps having to do with primordial fields!

\section{Magnetic field evolution in the early Universe}

Primordial magnetogenesis scenarios generally lead to fields
which are Gaussian random, characterized by a spectrum $M(k)$ (see below).
This spectrum is normalized by giving the field strength
$B_0$, at some fiducial scale and as measured at the present epoch, 
assuming it decreases with expansion as $B=B_0/a^2(t)$, where $a(t)$ 
is the expansion factor.
We take $B_0$ to be a free parameter since the predictions
for its value from magnetogenesis theories are highly
model and parameter sensitive.
Note that the magnetic and radiation energy densities both scale with
expansion as $1/a^4$. So we can characterize the magnetic field effect
by the ratio $B_0^2/(8\pi\rho_{\gamma 0}) \sim 10^{-7} B_{-9}^2$
where $\rho_{\gamma 0}$ is the present-day energy density in radiation, and 
$B_{-9} = B_0/(10^{-9} G)$.
Magnetic stresses are therefore small compared to
the radiation pressure for nG fields. 
The scalar, vector and tensor parts of the
perturbed stress tensor associated with magnetic fields
lead to corresponding metric perturbations.
Further the compressible part of the Lorentz force leads to
compressible (scalar) fluid velocity and associated density
perturbations, while its vortical part leads to
vortical (vector) fluid velocity perturbation.
The magnetically induced compressible fluid perturbations,
for nG fields, are highly subdominant compared to 
those due to the inflationary scalar modes. 
More important in our context will be the Alfv\'en mode driven by
the rotational component of the Lorentz force, especially 
since they decay without such magnetic driving. 
We recall below some features of their
evolution (Jedamzik, Katalinic \& Olinto 1998 (JKO); 
Subramanian \& Barrow 1998 (SB98a)).

The Alfv\'en mode oscillates negligibly on Mpc Scales by recombination,
with the phase of its oscillation 
$\chi = kV_A\eta\sim 10^{-2} B_{-9} (k/0.2h {\rm Mpc}^{-1}) (\eta/\eta_*)$.
Here the Alfv\'en velocity is $V_{A} \sim 3.8 \times 10^{-4} c B_{-9}$, $k$
the comoving wavenumber, $\eta$ the conformal time with $\eta_*$ 
its value at recombination. Unlike the compressional mode, which gets
strongly damped below the Silk scale $L_{S}$
due to radiative viscosity 
(Silk 1967), the Alfv\'en mode behaves like an over-damped oscillator.
Note that for an over-damped oscillator there is one normal mode which
is strongly damped and another one where the velocity starts from
zero and freezes at the terminal velocity till the damping becomes
weak at a latter epoch. The net result is that  
the Alfv\'en mode survives Silk damping for scales 
bigger than $L_A \sim (V_A/c) L_{S} \ll L_{S}$, the
canonical Silk damping scale (JKO; SB98a). 
The resulting baryon velocity is potentially detectable, 
since due to the Doppler effect, CMB temperature anisotropies
$\Delta T/T \sim V/c \sim f_d \chi V_A/c \sim 10^{-3} f_d \chi (B_{-9}/3) $
are induced. Here $f_d < 1$ takes into account
possible damping effects due the thickness of the last scattering
surface. One sees that for nG fields, significant
temperature anisotropies with $\Delta T/T \sim 10^{-6}$
can result, even for $f_d \chi \sim 10^{-3}$.
Detailed computations bear out this simple estimate and 
show that the signal peaks on arc minute scales, 
$l \sim 10^3$ (corresponding to the angle subtended by the
Silk scale at recombination), and because $L_A \ll L_{Silk}$, it 
is significant even below the Silk scale.

After recombination, when radiation decouples from the baryons,
the cosmic pressure drops by a large factor of order the photon to baryon
ratio, $n_{\gamma}/n_b \gg 1$.
The surviving tangled magnetic fields can now drive strong
compressible motions and seed density fluctuations 
(Wassermann 1978; SB98a; Sethi 2003), which
could well be very important in forming the first
structures and leading to an early re-ionization 
of the Universe (Sethi \& Subramanian 2005).

\section{CMB signals from tangled magnetic fields}

CMB anisotropies in general arise in two ways. Firstly,
spatial inhomogeneities around the surface
of last scattering of the CMB lead to the  
'primary' anisotropies in the CMB temperature as seen
at present epoch. Furthermore, variations in 
intervening gravitational and scattering effects, 
which influence the CMB photons as they come to us from 
the last scattering surface, can lead 
to additional secondary anisotropies
(cf. Subramanian 2005 for a review).
The CMB temperature anisotropies $\Delta T(\theta,\phi)/T$ are
expanded in spherical harmonics, 
\begin{equation}
{\Delta T\over T}(\theta,\phi)
= \sum_{lm} a_{lm} Y_{lm}(\theta,\phi) ;
\quad a_{lm}^* = (-1)^m a_{l-m},
\end{equation}
and expressed
in terms of their angular power spectrum $C_l$ where
the ensemble average 
$\bra{a_{lm}a_{l'm'}^*} = C_l\delta_{ll'}\delta_{mm'}$.
The mean square temperature fractional anisotropy is given by
\[
\frac{\bra{(\Delta T)^2}}{T^2} = \sum_l C_l {2l +1\over 4\pi} \approx
\int \frac{l(l+1)C_l}{2\pi} \ d \ln l
\]
with the last approximate equality valid for large $l$. So
$(l(l+1)C_l)/2\pi$ measures the power in the temperature
anisotropies per logarithmic interval in $l$ space.
(This combination is used because
scale-invariant potential perturbations generate anisotropies,
which at large scales (small $l$) have a nearly constant
$l(l+1)C_l$). A convenient characterization
 of the scale-dependent temperature anisotropy
is $\Delta T(l) = T [l(l+1)C_l)/2\pi]^{1/2}$. This is plotted
in Figure 1, as a dashed-triple-dotted line for a standard
$\Lambda$CDM model.

Primordial magnetic fields induce a variety of additional
signals on the CMB. An uniform field would for example select out
a special direction, lead to anisotropic expansion around this direction,
hence leading to a quadrupole anisotropy. The 
degree of isotropy of the CMB then implies a limit
of several nG on such a field 
(Barrow, Ferreira \& Silk 1997). Comparable limits 
may obtain, at least for the uniform component,
from upper limits to the IGM 
Faraday rotation of high redshift quasars 
(cf. Blasi et al. 1999).
For inhomogeneous, tangled primordial fields, 
the spatial inhomogeneities around the surface
of last scattering, due to magnetically induced 
perturbations (scalar, vector and tensor), 
lead to both large and small 
angular scale anisotropies in the CMB temperature 
and polarization.

For magnetically induced scalar perturbations,
the question of the initial conditions 
has only been analyzed in some detail recently in a paper
by Giovannini (2004), and detailed predictions of the CMB
anisotropies, for general initial conditions, are yet to be worked out.
(For some approximate treatments of the magnetosonic scalar modes
see Adams et al. 1996; Koh \& Lee 2000; 
Yamazaki, Ichiki \& Kajino 2005).
Further, scalar modes are dominated by the standard
inflationary scalar perturbation and are also damped by radiative
viscosity below the Silk scale.

Much more work has been done on the vector modes 
(Subramanian \& Barrow 1998 (SB98b), 2002 (SB02);
Seshadri \& Subramanian 2001 (SS01); 
Mack, Kahniashvilli \& Kosowsky 2002; 
Subramanian, Seshadri \& Barrow 2003 (SSB03); Lewis 2004). 
They typically lead to a temperature anisotropy 
$\Delta T \sim 5 \mu {\rm K}\, (B_{-9}/3)^2$ at $l \sim 1000$ and above (see below).
A comparable signal
arises at large angular scales, $l < 100$, 
due to gravitational wave perturbations (tensors).
All modes lead to much smaller polarization signals. 
The tensor and vector components in particular also lead to 
B-type polarization, which can help distinguish magnetic field
induced signals from those due to inflationary scalar modes.

In addition, the presence of tangled magnetic fields 
in the intergalactic medium
can cause Faraday rotation of the polarized component
of the CMB, leading to the generation of new B-type
signals from the inflationary E-mode signal. Their damping
in the pre-recombination era can lead to spectral distortions
of the CMB (Jedamzik, Katalinic \& Olinto 2000), 
while their damping in the post-recombination
era can change the ionization and thermal history of the Universe.
A potentially important consequence which we discuss below
is the magnetic field induced structure formation, which
may be relevant to explain the early re-ionization implied by
the WMAP data. We discuss some of these effects further below,
focusing in more detail on vector modes and post recombination effects,
where we have been directly involved in the computation of the
magnetic signals.

\subsection{Vector modes}

On galactic scales and above, the induced velocity due to the
Lorentz forces is generally so small
that it does not lead to any appreciable distortion of the initial field
(JKO; SB98a). Hence,
the magnetic field simply redshifts away as ${\bf B}(%
{\bf x},t)={\bf b}_{0}({\bf x})/a^{2}$. The Lorentz force associated with
the tangled field ${\bf F}_{L}={\bf F}/(4\pi a^{5})$, with 
${\bf F} = ({\bf \nabla }\times {\bf b}%
_{0})\times {\bf b}_{0}$, pushes the fluid to create
rotational velocity perturbations. These can be estimated 
by using the Navier-Stokes equation for the baryon-photon fluid 
in the expanding Universe,
\begin{equation}
\left( {\frac{4}{3}}\rho _{\gamma }+\rho _{b}\right) {\frac{\partial 
v_{i}}{\partial t}}+\left[ {\frac{\rho _{b}}{a}}{\frac{da}{dt}}+{\frac{
k^{2}\eta }{a^{2}}}\right] v_{i}={\frac{P_{ij}{\hat F}_{j}}{4\pi a^{5}}}.
\label{euler}
\end{equation}
Here, $\rho _{\gamma }$ is the photon density, $\rho _{b}$ the baryon
density, and $\eta =(4/15)\rho _{\gamma }l_{\gamma }$ the shear viscosity
coefficient associated with the damping due to photons, where 
$l_\gamma$ is the photon mean free path.
The projection tensor, $P_{ij}({\bf k})=[\delta _{ij}-k_{i}k_{j}/k^{2}]$
projects ${\bf {\hat F} }$, the Fourier component of  
${\bf F}$ onto its transverse components 
perpendicular to ${\bf k}$.

One can solve Eq.~\ref{euler} in two asymptotic limits, which allows
semi-analytic estimates of the signals. For scales
larger than the Silk scale, $kL_{S} < 1$, the radiative viscous damping
can be neglected to get 
$v_i = 3 P_{ij}{\hat F}_{j} D(\eta)/(16\pi\rho_0) $, where
$D(\eta) =\eta/(1 +R_*)$. Since $\vert {\hat F}_{ij}\vert \sim k V_A^2$,
we get for large scales $v/c\sim \chi V_A/c$, as
in our earlier simple estimate.
For scales smaller than the Silk scale, $kL_{S} > 1$, 
one assumes a terminal velocity approximation to balance 
viscous damping with the driving due to the Lorentz force.
This then leads to $D(\eta) \sim 5/ck^2L_\gamma$,
and $v/c \sim (5/kL_\gamma)V_A^2/c^2$, where
$L_\gamma$ is the co-moving photon mean free path.

The angular power spectrum $C_l$ of CMB anisotropies due to
rotational velocity perturbations is given by (Hu \& White 1997 (HW97))
\begin{eqnarray}
C_{l} &=& 4\pi \int_{0}^{\infty }{\frac{k^{2}dk}{2\pi ^{2}}}\quad {\frac{%
l(l+1)}{2}} \nonumber \\
&\times& \quad  
<|\int_{0}^{\eta_{0}}d\eta g(\eta_{0},\eta )v(k,\eta ){\frac{%
j_{l}(k(\eta_{0}-\eta ))}{k(\eta_{0}-\eta )}}|^{2}> 
\label{deldef}
\end{eqnarray}
Here $v(k,\eta )$ is the magnitude of the rotational component
of the fluid velocity $v_{i}$ in Fourier space,
and $\eta_0$ the present value of $\eta$.
The 'visibility function' $g(\eta_{0},\eta )$ determines the
probability that a photon reaches us at epoch $\eta_{0}$ if it
was last scattered at the epoch $\eta $.
So it weighs the contribution at any conformal time $\eta$ by the probability
of last scattering from that epoch.
We have shown as a solid line in Fig.~\ref{vis}
the visibility function for a standard $\Lambda$CDM model.
The spherical Bessel function of order $l$, the $j_{l}(z)$ term,
projects variations in space, at the conformal time $\eta$ around
the last scattering epoch, to angular (or $l$) anisotropies
at the present epoch. 
These spherical Bessel
functions generally peak around $k(\eta_0 -\eta) \approx l$.
The multipoles $l$ are then probing
generally spatial scales with wavenumber $k \sim l/(\eta_0 -\eta)$
at around last scattering.  

We assume that ${\bf b}_0$ is a Gaussian random field. 
Its Fourier components satisfy
 $<b_{i}({\bf k})b^*_{j}({\bf q})>
=\delta _{{\bf k},{\bf q}}P_{ij}({\bf k})M(k)$,
where the magnetic power spectrum is normalized 
using a top hat filter in k-space, and taking 
$k^{3}M(k)/(2\pi ^{2}) = (B_{0}^{2}/2)(n+3)(k/k_{G})^{3+n}$ with
$n > -3$. We generally normalize the field at $k_G = 1$ h Mpc$^{-1}$.
The spectrum is cut-off at $k_c$, determined
by dissipative processes.

An analytic estimate of the temperature anisotropy,
for $kL_S < 1$ is then (SB98b; SB02) 
\begin{equation}
\Delta T_{B}(l) 
\approx 5.8\mu K\left( {\frac{B_{-9}}{3}}\right) ^{2}\left( {\frac{l}{%
500}}\right) I({\frac{l}{R_{\ast }}}),
\label{an1}
\end{equation}
whereas for scales smaller than Silk scales with $kL_S > 1$, 
\begin{equation}
\Delta T_{B}(l) 
\approx 13.0\mu K \left( {\frac{B_{-9}}{3}}\right) ^{2}\left( {\frac{l}{%
2000}}\right) ^{-3/2}I({\frac{l}{R_{\ast }}}).
\label{an2}
\end{equation}
Here $I(k)$ is a mode coupling integral
\begin{eqnarray}
I^{2}(k)&=& {\frac{8}{3}}(n+3)({\frac{k}{k_{G}}})^{6+2n}; \quad n < -3/2
\nonumber \\
&=& {\frac{28}{15}}{\frac{(n+3)^{2}}{(3+2n)}}({\frac{k}{k_{G}}})^{3}({%
\frac{k_{c}}{k_{G}}})^{3+2n}; \quad n>-3/2.
\nonumber
\end{eqnarray}
Note that for $n< -3/2$, $I$ is independent of $k_c$.
For a nearly scale-invariant spectrum, say with $n=-2.9$, we get 
$\Delta T(l)\sim 4.7\mu {\rm K}\,(l/1000)^{1.1}$ for scales larger than the Silk
scale, and $\Delta T(l)\sim 5.6\mu {\rm K}\,(l/2000)^{-1.4}$ for scales smaller
than $L_{S}$ but larger than $L_{\gamma }$. Larger signals will be expected
for steeper spectra, $n>-2.9$ at the higher $l$ end.

One can also do a similar calculation for the expected CMB 
polarization anisotropy (SS01; SSB03). 
Note that polarization of the CMB arises 
due to Thomson scattering of radiation
from free electrons and is sourced by the quadrupole component of
the CMB anisotropy. 
For vector perturbations, the B-type contribution dominates
the polarization anisotropy (HW97), unlike for 
inflationary scalar modes. Further, as mentioned earlier,
the vector mode signals can also be important
below the Silk damping scale ($l\ge 10^3$).

\begin{figure}[t!]\centering
\includegraphics[width=0.5\textwidth,angle = 0]{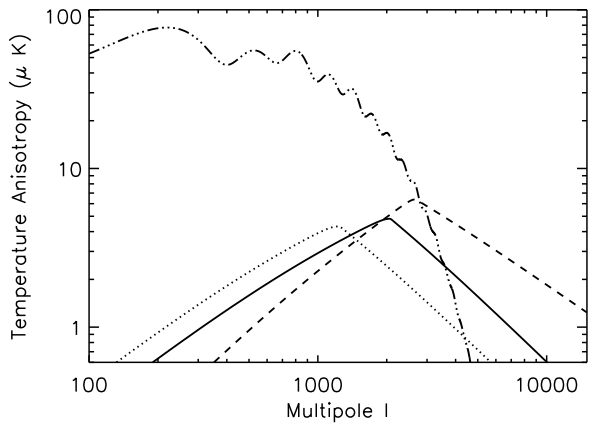}
\caption{$\Delta T$ versus $l$ predictions for different cosmological models
and $M(k)\propto k^{n}$, for $B_{-9}=3$. The bold
solid line is for a canonical flat,  $\Lambda $-dominated
model, with $\Omega _{\Lambda }=0.7$, $\Omega _{m}=0.3$, $\Omega
_{b}h^{2}=0.02$, $h=0.7$ and almost scale-invariant spectrum $n=-2.9$. The
dotted curve (....) obtains when one changes to $\Omega _{m}=1$ and $\Omega
_{\Lambda }=0$ model. The dashed line is
for the  $\Lambda $-dominated model with a larger baryon density $\Omega
_{b}h^{2}=0.03$, and a larger $n=-2.5$. We also show for qualitative
comparison (dashed-triple dotted curve),
the temperature anisotropy in a
'standard'  $\Lambda$CDM model, computed using CMBFAST
(Seljak \& Zaldarriaga 1996) with cosmological parameters as for the
first model described above. 
(Adapted from SB02 and SSB03)}
\label{vec}
\end{figure}

\begin{figure}[t!]\centering
\includegraphics[width=0.5\textwidth,angle = 0]{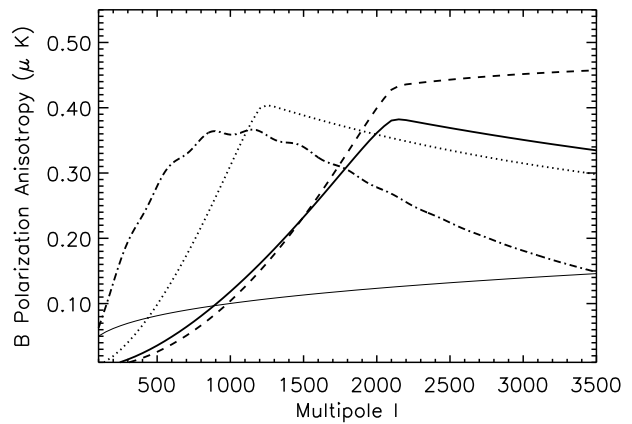}
\caption{$\Delta T_P^{BB}$ versus $l$ predictions for different cosmological
models and magnetic power spectrum $M(k)\propto k^{n}$, for $B_{-9}=3$.
The bold solid line is for a standard flat,  
$\Lambda $-dominated model, with $\Omega _{\Lambda }=0.73$,
$\Omega _{m}=0.27$, $\Omega_{b}h^{2}=0.0224$, $h=0.71$ and almost scale
invariant spectrum $n=-2.9$. The dashed curve obtains
when one changes to $n=-2.5$.
The dotted curve gives results for a $\Omega _{m}=1$ and
$\Omega_{\Lambda }=0$ model, with $n=-2.9$.
We also show for qualitative comparison
(dashed-dotted curve), the B-type polarization anisotropy due to
gravitational lensing, in the canonical $\Lambda$CDM model, computed
using CMBFAST (Seljak \& Zaldarriaga 1996).
The signal due to magnetic
tangles dominate for $l$ larger than about $1000$. Finally, the thin
solid line gives the expected galactic foreground contribution
estimated by Prunet et al. (1998), which is also smaller
than the predicted signals.
(Adapted from SSB03)}\label{vecpol}
\end{figure}

We show in Figs.~\ref{vec} and \ref{vecpol} the temperature
and polarization anisotropy for the magnetic field induced
vector modes obtained by evaluating the $\eta $ and $k$ integrals in 
Eq. (\ref{deldef}) numerically. We retain the analytic 
approximations to $I(k)$
and $v_{i}(k)$. These curves show the build up of power in temperature and 
B-type polarization due to vortical perturbations
from tangled magnetic fields which survive Silk damping at high
$l \sim 1000-3000$. The eventual slow decline is due to the
damping by photon viscosity, 
which is only a mild decline as the magnetically sourced vortical 
mode is over damped.
By contrast, in the absence of magnetic tangles there is a
sharp cut-off due to Silk damping. 
Our numerical results are consistent analytic estimates
given in Eqs.~(\ref{an1}) and (\ref{an2}).
A scale-invariant spectrum of tangled fields with
$B_{0}=3\times 10^{-9}$ Gauss, produces 
reduces temperature anisotropies at the $5\mu$K level 
and B-type polarization anisotropies  
$\Delta T_B \sim 0.3-0.4 \mu$K between
$l\sim 1000-3000$. Larger signals result for steeper spectra
with $ n > -3$.  Note that the anisotropies in hot or cold
spots could be several times larger, because the non-linear dependence of $%
C_{l}$ on $M(k)$ will imply non-Gaussian statistics for the anisotropies.

These magnetically induced signals could therefore contribute
to the excess power detected by the CBI experiment.
A distinguishing feature, compared to say the Sunyaev-Zeldovich 
(SZ) effect (Zeldovich \& Sunyaev 1969) canonically
invoked to explain the CBI excess, will be provided by
the B-type polarization signals. Note that if the observed excess power
in the CBI experiment arises from the SZ effect, this is not
expected to be strongly polarized. The spectral dependence
of the SZ signal could also help to distinguish it from
any magnetically induced signals, which are expected to be
frequency independent, at least at high frequencies. 

Recently, Lewis (2004) has done a detailed numerical computation of
the vector mode signal and finds qualitatively similar
results to the semi-analytical estimates given above.

\subsection{Tensor modes}

Tangled magnetic fields also produce
anisotropies on large angular scales, or small $l$, 
dominated by tensor metric
perturbations induced by anisotropic magnetic stresses (Durrer, Ferreira \&
Kahniashvili 2000; Mack, Kahniashvili \& Kosowsky 2002; 
Caprini \& Durrer 2002; Giovannini 2005b).
The tensor metric perturbation $h_{ij}$ 
obeys the equation
\[
h_{ij}'' +2 {\cal H} h_{ij}' -\nabla^2h_{ij} 
= -16\pi G a^2 \delta T_{ij}^{TT}
\]
where ${\cal H}=a'/a$,
a prime denotes derivative with respect to the conformal
time, and $\delta T_{ij}^{TT}$ is the transverse, traceless component
of the energy momentum tensor (due to the magnetic field).
The resulting CMB anisotropy is then computed using
\[
(\Delta T/T) = -\frac{1}{2} \int_{\eta_i}^{\eta_0}
h_{ij}' n^i n^j d\eta
\]
where $n^i$ is a unit vector along the line of sight, and
prime denotes a conformal time derivative.
Using the formalism
described in these papers, we estimate a tensor contribution at small $l<100$
of $\Delta T \sim 7 (B_{-9}/3)^2 (l/100)^{0.1} \mu{\rm K}$, for $n=-2.9$.
(One has to account for neutrino anisotropic stress compensation
(Lewis 2004) after neutrino decoupling).
Since we have to add this power to the standard power
produced by inflationary scalar perturbations in quadrature, a tangled field
with $B_{-9}\sim 3$ will produce of order a few to 10 percent
perturbation to the power in the standard CMB anisotropy at large angular 
scales.
So if they are indeed detected at large $l$, below the Silk damping scale,
one will also have to consider their effects seriously at
large angular scales, especially in cosmological parameter estimation.  
The tensor mode also contributes to the B-type polarization
anisotropy at large angular scales ($l <100$ or so), with
$\Delta T_B < 0.1 \mu $K for $B_{-9} < 3$.
The production of gravitational waves has been used in an indirect
manner by Caprini \& Durrer (2002) to set strong upper limits on  
$B_0$ for spectra with $n > -2.5$ or so.

\subsection{Faraday rotation due to primordial fields}

Another interesting effect of primordial fields is the
the Faraday rotation it induces on the polarization of
the CMB (Kosowsky \& Loeb 1996; Kosowsky et al. 2005;
Campanelli et al. 2005). 
The rotation angle is about
\begin{eqnarray}
\Delta\Phi &=& \lambda_0^2 \ (3/e)
\int_0^{\eta_0} d\eta' g(\eta_0, \eta') {\bf n}
\cdot{\bf B}_0[{\bf n}(\eta'-\eta_0)]
\nonumber \\
&\approx& 1.6^o B_{-9} (\nu/30 {\rm GHz})^{-2}
\nonumber
\end{eqnarray}
where $\lambda_0$ is the wavelength of observation. 
So this effect is important only at low frequencies,
and here it can lead to the
generation of B-mode polarization from the Faraday
rotation of the inflationary
E-mode. From the work of Kosowsky et al. (2005) one 
can estimate a B-mode signal 
$\Delta T_B \sim 0.4 (B_{-9}/3)\, (\nu/30 {\rm GHz})^{-2} \mu {\rm K}$,
for $n=-2$, at $l \sim 10^4$.
The signals are smaller at smaller $n$.
The Faraday rotation signal can be distinguished
from the B-mode polarization generated by say
vector modes, or gravitational lensing, because of
their frequency dependence ($\nu^{-2}$).

\section{Post recombination blues}

\begin{figure}[t!]\centering
\includegraphics[width=0.35\textwidth,angle = -90.0]{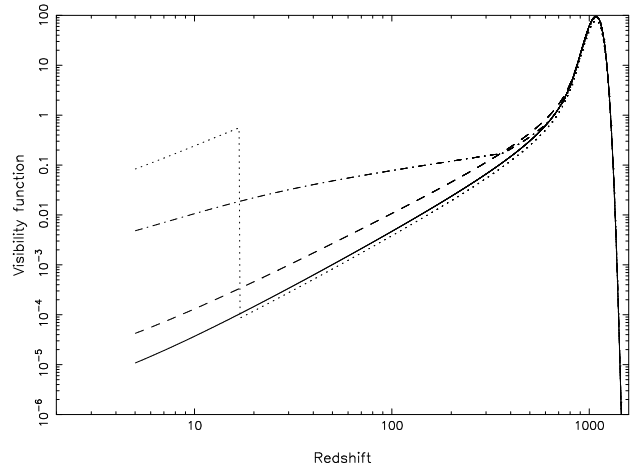}
\caption{Visibility function for different models. The
solid and the dotted  curves are for the standard recombination
and a model in which the Universe re-ionizes at $z =17$, respectively.
The dashed curve corresponds  to a  decaying turbulence model with 
$B_0 = 3 \times 10^{-9} \, \rm G$.
The dot-dashed  curve corresponds to the
ambipolar diffusion case with $B_0 = 3 \times 10^{-9} \, \rm G$
and $n = -2.8$. (Adapted from SS05)}
\label{vis}
\end{figure}

After recombination the Universe
became mostly neutral, resulting also in a sharp drop in the
radiative viscosity.  Primordial magnetic fields can then dissipate
their energy into the intergalactic medium (IGM) via ambipolar diffusion and,
for small enough scales, by generating decaying MHD turbulence.
These processes can significantly modify the thermal and ionization history of
the post-recombination Universe. We show in Fig.~\ref{vis} 
the modified visibility function 
due to the gradual re-ionization by ambipolar damping and 
turbulence decay from the work of Sethi \& Subramanian (2005).

These dissipative processes, for $B_{-9} \sim 3$, 
can give rise to Thomson scattering optical depths $\tau \ga 0.1$,
although not in the range of redshifts needed to explain the
recent WMAP polarization observations (the T-E cross correlation
seen at low $l$). 
However, future CMB probes like PLANCK
can potentially detect the modified CMB anisotropy signal from
such partial re-ionization (Kaplinghat et al. 2003).
This can be used to detect or further constrain small-scale
primordial fields.

\begin{figure}
\includegraphics[width=0.35\textwidth,angle = -90.0]{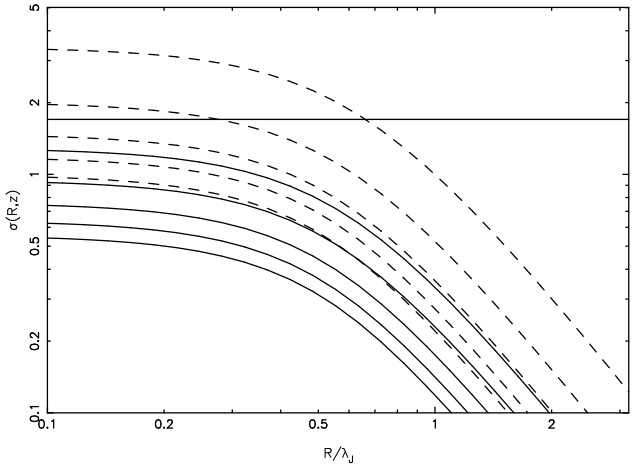}
\caption{The mass dispersion $\sigma(R,z)$ is shown for two
models with nearly scale free  magnetic field power spectra. 
The solid and dashed curves correspond to $n = -2.9$ and 
$n = -2.8$, respectively.
Different curves, from top to
bottom,  correspond to redshifts $z = \{10,15,20,25,30\}$, respectively.
The horizontal line corresponds to $\sigma = 1.68$. 
(Adapted from SS05)}
\label{struc}
\end{figure}

Potentially more exciting is the
possibility that primordial fields could  induce the formation
of subgalactic structures for $z \ga 15$.
We show in Fig.~\ref{struc} the mass dispersion $\sigma(R,z)$ for two
models with nearly scale-free magnetic field power spectra, as computed by
SS05. When $R$ is normalized to the magnetic Jeans scale, $\lambda_J$,
it turns out that $\sigma$ depends only on the ratio
$R/\lambda_J$ and not explicitly on the strength of the field. 
This interesting feature arises because density fluctuations are
generated by the divergence of ${\bf F}_L$ and so are $\propto k^2B_0^2$.
Since the magnetic Jeans scale $\lambda_J \propto k_J^{-1} \propto B_0$,
the magnetic field dependence cancels out in $\sigma$ when
scales are expressed in terms of $R/\lambda_J$ (see SS05 for details).
Structures collapse when $\sigma > 1$ (for the 
the spherical top hat model when $\sigma = 1.68$).
For the nearly scale-free power law models, even typical 
structures at the magnetic Jeans scale collapse at 
high redshifts in the range between $10$ to $20$.

The mass of these objects does 
depend on the magnetic field strength
smoothed to the Jeans scale, and lie in the range
$10^9 M_\odot$ to $3 \times 10^{10} M_\odot$
for $B_0 \sim 10^{-9} \, \rm G$ to $B_0 \sim 3 \times 10^{-9} \, \rm G$.
An even smaller field could have a major impact, provided the 
collapsing structures have a mass larger than the thermal Jeans mass.
For example, a field as small as $B=0.1$~nG can induce a 
$10^6 M_\odot$ dwarf galaxy collapse
at high z ($z > 15$), causing early enough re-ionization 
to explain the T-E cross correlation 
peak observed by WMAP. More detailed work on this aspect is underway.

\section{Conclusions}

We briefly reviewed some of the possible ways one could
detect/constrain primordial magnetic fields using the CMB
anisotropies and polarization. This endeavor would be even more
fruitful if there were a compelling mechanism for primordial
magnetognesis, which also produced strong enough ($B_{-9} \sim 1$) 
and coherent enough (ordered on Mpc scales) fields. 
At this juncture, theoretical predictions are
highly parameter dependent, and so we have taken a more pragmatic approach of
assuming that such a field could be generated in the early Universe
and asking what it would imply for the CMB and structure formation
in general. 

For a field of $B \sim 3$~nG and a nearly scale-invariant spectrum
one predicts CMB temperature anisotropies with a $\Delta T \sim 5 \mu$K,
at $l< 100$ and $l> 1000$ and polarization anisotropies with 
$\Delta T_P \sim 0.4 \mu$K at $l > 1000$. 
Especially interesting is that the vector modes induced
by primordial fields can contribute significantly below the Silk
scale, where the conventional scalar modes are exponentially damped.
Further, the magnetically induced signal at small angular scales
will be dominated by B-mode polarization.
There do exist intriguing results from the CBI experiment for the presence
of a temperature excess at small angular scales. Some part of this
excess could arise due to the influence of primordial fields. 
This excess is conventionally explained as arising due to the SZ effect, 
but the power in density fluctuations 
on cluster scales (conventionally measured by $\sigma_8$), 
has to be pushed to be in the upper range of values ($\sigma_8 \sim 1$), 
allowed by current CMB and large-scale structure data. 
Since the SZ signal is frequency dependent, a crucial test would 
be to compare CMB observations at different frequencies.

Also if B-type polarization at these scales is detected
this could be a good indicator of the magnetic field effects.
Clearly it will be important to make further observations
at small angular scales, especially at different frequencies.
It is also important to study the statistics
of the CMB anisotropies, since the magnetically induced
signals are predicted to be strongly non-Gaussian.
If magnetic fields have helicity, this would induce further interesting
effects, which have been reviewed in this meeting 
(Kahniashvilli 2006 and references therein).

We have also emphasized another interesting consequence of the
existence of primordial magnetic fields, which indirectly affects
the CMB anisotropies. Due to their presence the first collapsed objects
of dwarf galaxy masses and smaller can form at high $z > 15$,
even for $B \sim 0.1$~nG. This can potentially lead to the early re-ionization
indicated by the present WMAP polarization data in a very natural manner.

\acknowledgements
Much of the work reviewed here are in joint papers 
with John Barrow, T. R. Seshadri and Shiv Sethi. I thank them
for very enjoyable collaborations, Dmitri Sokoloff for useful 
comments, and the SKA Project Office for partial support.

\end{document}